\documentclass[sigconf,authorversion,nonacm]{acmart}
\usepackage{graphicx}
\usepackage{url}
\usepackage{xcolor}
\usepackage{tikz}
\usepackage{multirow}
\usepackage{enumitem}
\usepackage[autostyle]{csquotes}
\usepackage[misc,geometry]{ifsym}
\usepackage{placeins}

\newcommand*\ce[1][1ex]{\tikz\draw (0,0) circle (0.8ex);} %empty circle
\newcommand*\ch[1][1ex]{%%half circle
  \begin{tikzpicture}
  \draw[fill] (0,0)-- (90:0.8ex) arc (90:270:0.8ex) -- cycle ;
  \draw (0,0) circle (0.8ex);
  \end{tikzpicture}} %
\newcommand*\cf[1][1ex]{\tikz\fill (0,0) circle (0.8ex);} %%full circle
\newcommand\blfootnote[1]{%
  \begingroup
  \renewcommand\thefootnote{}\footnote{#1}%
  \addtocounter{footnote}{-1}%
  \endgroup
}

\AtBeginDocument{%
  }

\begin{document}
\title[short title for page header]{full title}
\title[Challenges and Opportunities in Securing Smartphones Against Zero-Click Attacks]{Experience Report on the Challenges and Opportunities in Securing Smartphones Against Zero-Click Attacks}
\author{Narmeen Shafqat$^\ast$, Cem Topcuoglu$^\ast$, Engin Kirda, Aanjhan Ranganathan} 
    \affiliation{ 
      \institution{Northeastern University} \city{Boston} \state{MA} \country{USA}
    }
    \email{{shafqat.n, topcuoglu.c, e.kirda, aanjhan}@northeastern.edu}

\renewcommand{\shortauthors}{Shafqat et al.}

\begin{abstract}

Zero-click attacks require no user interaction and typically exploit zero-day (i.e., unpatched) vulnerabilities in instant chat applications (such as WhatsApp and iMessage) to gain root access to the victim's smartphone and exfiltrate sensitive data. In this paper, we report our experiences in attempting to secure smartphones against zero-click attacks. We approached the problem by first enumerating several properties we believed were necessary to prevent zero-click attacks against smartphones. Then, we created a security design that satisfies all the identified properties, and attempted to build it using off-the-shelf components. Our key idea was to shift the attack surface from the user's smartphone to a sandboxed virtual smartphone ecosystem where each chat application runs in isolation. Our performance and usability evaluations of the system we built highlighted several shortcomings and the fundamental challenges in securing modern smartphones against zero-click attacks. In this experience report, we discuss the lessons we learned, and share insights on the missing components necessary to achieve foolproof security against zero-click attacks for modern mobile devices.  

\end{abstract}

\maketitle
\blfootnote{$^\ast$ Equal Contribution.}

\section{Introduction}

\begin{figure}[t]
\begin{center}
\includegraphics[width=\linewidth]{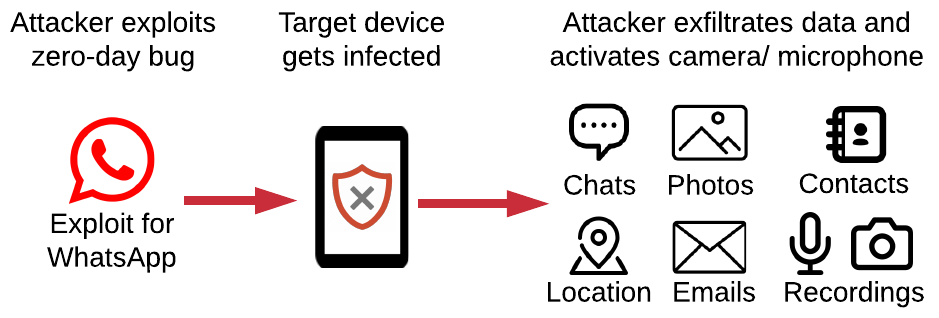}
\caption{Attacker can exploit zero-click zero-day bug in a chat application (e.g. WhatsApp) to gain root access to the target device and exfiltrate sensitive information.}
\label{fig:problem}
\end{center}
\end{figure}

The increasing use of smartphones for communications, such as banking and social networking, has made them an attractive target for cyber criminals.  
These malicious actors used social engineering to lure victims into clicking a malicious link or pressing a button, thereby causing the malware to execute, proliferate and compromise the victim's smartphone successfully.  
However, the interaction requirement has made it difficult for the attacker to compromise the technically-savvy targets.
Cyber criminals are now using zero-click exploits that do not require interaction from the user and abuse zero-day (i.e., unpatched) vulnerabilities in smart applications, typically instant messaging applications, to control the victim's smartphone.  
Zero-click attacks are stealthy in nature, and avoid persistence to escape detection by anti-malware utilities and forensic tools~\cite{forensics}.
This makes it extremely difficult to detect and prevent zero-click attacks.
Recently, it was discovered that a sophisticated spyware, named \textit{Pegasus}~\cite{pegasus}, had used zero-click exploits to spy on renowned journalists, human rights defenders, political dissidents, business executives, and lawyers around the world for several years.  
With a successful zero-click attack, the attacker can gather sensitive data (such as messages, contacts, photos, files, emails, and usage history), log the victim's keystrokes (which can further leak passwords, security tokens, and credit card information), determine the victim's location, and even remotely activate the device's camera and microphone at any time for real-time surveillance (Figure~\ref{fig:problem}). 
In short, zero-click attacks completely invade the victim's privacy and threaten personal security.  
As a matter of fact, the attacker only requires the victim's phone number to send the malicious zero-day exploit; hence acquiring access to the victim's contact list also endangers the smartphones of many more users. 

Given that the zero-click attacks largely exploit zero-day vulnerabilities, existing measures such as mobile anti-virus or malware detection systems have completely failed to thwart zero-click attacks.  
%So far, there is no concrete safeguard against zero-click attacks, besides only recommendations by security experts to keep the mobile OS updated and remove unnecessary applications from the smartphone.  
Currently, Mobile Verification Toolkit (MVT)~\cite{mvt}, developed by Amnesty International Security Lab, is the only tool that can analyze device backup to find signs of potential zero-click compromise (provided such attack vectors are already known).
%Tech giant, Apple, is soon launching \textit{Lockdown mode} with the release of iOS 16, that aims to 
In the absence of any real-time counteractive measure against zero-click attacks, the user himself is responsible for securing the smartphone. 

A naive way of protecting one's privacy from zero-click exploits is not to use a smartphone or consider using a burner phone with a burner sim. 
However, such options limit the user's ability to even use the Internet for browsing.  
Alternatively, the user may resort to a secondary smartphone for running chat applications, but a zero-click exploit targeting a single chat application can also leverage cross-application vulnerabilities to compromise other chat applications on the smartphone.
Nevertheless, the user may rely on web or desktop clients instead of smart applications, e.g., for WhatsApp. However, this requires him to carry the laptop everywhere. Also, not all chat applications have web/ desktop interfaces; thus, even allowing just SMS messages to be delivered on the smartphone can risk the smartphone's security. 
In essence, there is an urgent need to design and develop security mechanisms that protect against zero-click attacks. 
More importantly, the security framework should primarily aim to significantly limit the loss of information in case of an exploit, as software bugs are here to stay. 

\begin{figure}[t]
\begin{center}
\includegraphics[width=\linewidth]{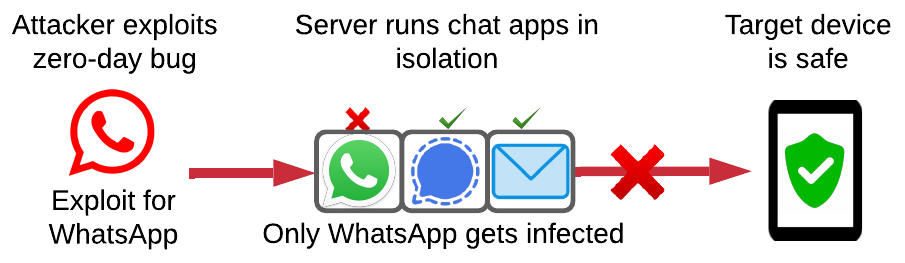}
\caption{Our attempt at building a Zero-Click Secure Architecture: The user accesses remote isolated chat applications via screen sharing, which confines zero-click exploit to the targeted application only, and safeguards other applications on the server as well as the target device.}
\label{fig:benefit}
\end{center}
\end{figure}

In this paper, we report on our experiences and lessons learned in our attempt to design and develop a security framework against zero-click attacks.  
Our goal was to provide high-risk individuals, such as investigative journalists, who reached out to us with a secure solution that allowed them to use their chat application of choice while guaranteeing maximum privacy in case of a surveillance attack.  
We started off by studying the zero-click attack landscape and enumerated several properties we believe are necessary to limit the impact (e.g., amount of information leaked) and lifespan of the attack (e.g., duration of the surveillance). 
For example, as zero-click exploits primarily leverage vulnerabilities within and across chat applications to compromise a device, it was essential to prohibit any direct delivery of messages to the user's smartphone.  
Our key intuition was that a remote access mechanism was needed such that the user could interact with the messages without worrying about any potential exploit that could compromise the user's smartphone. 
To this extent, we attempted to build a zero-click secure framework that shifted the attack surface from the user's device to a virtual smartphone ecosystem (remote server) that runs each chat application in a sandbox-like isolated environment (Figure~\ref{fig:benefit}). 
The motivation for our design was to answer the question: Can we build a zero-click secure framework using readily available off-the-shelf components? 

We practically implemented a proof-of-concept using commercial off-the-shelf (COTS) based software components by running each application in a containerized Android emulator and using Web Real-Time Communication (WebRTC) for screen sharing and remote interaction.  
To understand the feasibility and practicality of our solution, we conducted a user study that primarily evaluated usability and performance.  
Note that the user study had received exemption by the Institutional Review Board (IRB) under the IRB Category-3 Benign Behavioral Interventions.  
Our evaluations highlighted several shortcomings and fundamental challenges that exist in securing smartphones against zero-click attacks.  

Specifically, this paper makes the following contributions:
\begin{enumerate}
\item We enumerate several design requirements to protect smartphones against zero-click attacks. 
We assume that zero-day vulnerabilities will continue to exist, and therefore, our requirements are geared towards building a scalable and usable security architecture that strictly prevents zero-click exploits from compromising the user's device, and limits the impact and lifespan of a potential attack. 
\item We evaluate the performance and usability of the secure framework we built to attempt to solve the zero-click problem, and highlight the challenges while building such a solution using COTS components.
\item We distill five concrete lessons from our experience of attempting to build a zero-click secure architecture for smartphones. 
We discuss our experiences in finding a reliable, scalable and cost-effective sandboxing solution to run the mobile OS on a remote server 24/7, and our attempts to create a system that works over the Internet with minimal network lag and components that allow users to interact with the apps seamlessly with a high degree of interaction quality.

\end{enumerate}

Our aim in this work is not to present a solution against zero-click attacks that works perfectly in practice. 
Rather, through this work, based on our attempts (some of which were not very successful) to build such a system using standard software components, we aim to identify the challenges, limitations, and key research opportunities towards realizing a zero-click protection system for users. 
We hope our experiences will be useful to other researchers working to address zero-click attacks against mobile devices.

\section{Background and State-Of-The-Art} \label{background-motivation}
\subsection{Zero-click Exploits}

Zero-click exploits tend to leverage common instant messaging applications such as WhatsApp and iMessage that, by design, receive messages and calls from anyone who knows the user's phone number, including untrusted sources.  
The attacker gears a zero-click attack by sending a specially-crafted hidden text message, image, or voicemail to the target device via a wireless connection (Wi-Fi, cellular network, Bluetooth, or NFC).  
The injected malicious code then provokes a previously unknown security vulnerability in an installed application to gain root access to the target device.    

Zero-click exploits are not new.  Apple first discovered a zero-click vulnerability on the iPhone 5 with iOS 6 in 2012~\cite{apple2012}.  
With the increase in attack sophistication and complex computing technologies, zero-click attacks have become a pervasive threat to smartphone users. 
Samsung recently patched a zero-click vulnerability in Skia (Android's graphics library) that existed in all Samsung smartphones sold since 2014~\cite{androidmms}.
Zero-click attacks have gained significant attention since the recent discovery of Pegasus, the zero-click spyware used to surveil high-profile individuals and organizations around the world.

\subsubsection*{\textbf{Pegasus Spyware}} 
Pegasus is a highly sophisticated cyber-espionage spyware created by NSO Group Technologies Ltd, an Israeli cybersecurity company, with the claimed aim of tracking terrorist activities via untraceable commands~\cite{pegasus}.
However, several reports claim that nation-state actors have misused Pegasus spyware to track the activities of their opponents and critics. 

In 2016, Apple discovered three zero-click exploits in iOS that were used to infect and spy on targets for years~\cite{forensics}.
In 2017, Pegasus spyware abused a zero-day vulnerability in WhatsApp with a mere unanswered message or call that enabled the spyware to run as a background resource and compromise the target phone. 
The Pegasus variants identified in 2020 and 2021 used two zero-click iMessage exploits; Kismet and ForcedEntry.  
Although Kismet was successful only against iOS release 13.5.1, ForcedEntry was exploited in the wild and even bypassed the iOS BlastDoor utility, which was basically designed to prevent such spyware attacks.  
Surprisingly, the 2021's Pegasus variant utilized the same vulnerability chaining technique, \textit{Trident}, as the 2016's Pegasus variant, which was patched by Apple right after the attack, albeit ineffectively~\cite{lookout}.

A key trait of Pegasus (and, in general, zero-click spyware) is that it does not show erratic behaviour or leave indicators of compromise (IoC) such as slow performance or excessive battery drainage.  
The spyware primarily exists in the phone's temporary memory (RAM) and is intermittently removed.  
Even upon installation on the phone, the spyware does not leave any trace.  
This way, it effectively evades detection by anti-malware or endpoint security systems, thereby failing to inform the user about the potential privacy infringement and taking security
threats to the next level. 

\subsubsection*{\textbf{Other zero-click spyware:}} 
Zero-click exploits now have a thriving market.  
Reign, zero-click spyware developed by QuaDream, can also compromise iOS devices using an iMessage-based zero-day exploit~\cite{reign}.  
In addition, sources claim that the zero-click exploits developed by Paragon~\cite{paragon} and Cognyte~\cite{threemore} can reportedly target end-to-end encrypted messaging applications, particularly WhatsApp and Signal to compromise the smartphone.  

\subsection{Current State-Of-The-Art} \label{related-work}

The technical details regarding how a zero-click exploit stealthily compromises a smartphone without user interaction are largely unknown.  
For this reason, the existing proactive and reactive security systems, such as mobile anti-virus, intrusion detection and prevention systems that detect the presence of advanced mobile malware by analyzing parameters like requested permissions~\cite{permission}, API and system calls~\cite{apicall, droidcat}, network addresses~\cite{clickip}, resource consumption~\cite{resources}, etc., have completely failed to detect and prevent zero-click attacks.
Even the security experts have only put forward recommendations to keep the mobile OS and installed applications updated, which cannot be considered a panacea for secure communication. 

Over the years, Amnesty International Security Lab forensically investigated several Pegasus-infected smartphones. The lab publicly disclosed their forensic methodology~\cite{forensics} and released an automated tool, MVT~\cite{mvt}, that analyzes device backups to verify if the smartphone is infected with Pegasus spyware or not.  
However, MVT cannot detect and prevent zero-click attacks in real time or stop the leakage of sensitive information after the attack execution.  
  
In an attempt to guarantee protection against mercenary zero-click exploits, tech giant Apple recently rolled out an optional feature of \textit{Lockdown mode} with the release of iOS 16 in fall 2022~\cite{lockdown}. 
The Lockdown mode aims to strictly restrict certain features such as receiving texts from unknown numbers, displaying link previews in messages, wired connections with computers, etc.
The Lockdown mode was found to be overly prohibitive as the user cannot even download and view any attachment (other than images) on the phone. 
Since Lockdown mode disables custom fonts for websites to prevent execution of malicious JavaScript code, the website administrator can detect missing fonts on a device and indirectly know that the user is potentially a high-profile target. 
As websites already log the IP address of the visitors, the acquired information can be further used to fingerprint users or devices~\cite{lmfingerprinting}. 
While Apple users can benefit from this groundbreaking feature in the near future, Android users do not have any concrete defensive or preventive solution against zero-click attacks yet.

Unfortunately, most notable research works that claim to have high zero-day detection accuracy have been carried out on non-mobile OS~\cite{deep2020ids, network2020, network2021, cloud2015}, whereas the aforementioned zero-click exploits are carefully designed to abuse zero-day bugs in smart chat
applications.  
The few research works that focus on the detection of mobile-based zero-day malware, as discussed below, typically employ signature or anomaly-based detection schemes~\cite{survey2020}.

AndroSimilar~\cite{androsimilar2013} and DroidAnalytic~\cite{droid2013} specifically used advanced signature-based detection techniques to identify zero-day repackaged malware (i.e., unseen variants of the known mobile malware).
However, such approaches required exhaustive collection of malware samples and could not detect new zero-day vulnerabilities.  
Several studies also demonstrated the feasibility of detecting Android-based zero-day malware by analyzing features like API sequence call, opcode, permissions, etc., using machine learning approaches; Bayesian classification~\cite{bayesian2013, bernoulli2015}, and deep learning~\cite{deep2021, amin2020}.  
However, as zero-day malware evades anti-malware detection by hiding its malicious activity on detecting an isolated or sandboxed background, these detection approaches are extremely slow~\cite{mobile2019} and ineffective.

In contrast, existing anomaly-based detection approaches looked for deviations in smartphone activity from the normalized baselines~\cite{linear2013} to detect zero-day malware.  
RiskRanker~\cite{riskranker2012} and Andro-AutoPsy~\cite{andro2015} particularly observed if any application exhibits malicious activity, such as trying to launch a root exploit, whereas, DroidLight~\cite{droidlight2020} used one-class classification and probability distribution analysis to detect zero-day malware.  
However, where RiskRanker and DroidLight required intensive processing power, Andro-AutoPsy could not analyze malware that employed anti-malware and encryption techniques.  
In short, since the existing research works could not effectively detect zero-day malware on a mobile OS, they will also be ineffective against sophisticated zero-click exploits that use native libraries and advanced obfuscation techniques to hide their traces.

Owing to the lack of defensive solutions against zero-day attacks, we also investigated existing research studies on virtual smartphones~\cite{chen2010, svmp} that run Android OS on the cloud platform to run sensitive applications away from the user's device.  
However, such setups are now obsolete and do not support the latest versions of Android OS. 
To the best of our knowledge, currently, there is no holistic preventive or defensive solution that can be used as an immediate safeguard against zero-click attacks, particularly for Android users.
\section{Our Envisioned Secure Framework} \label{design}
\subsection{Threat Scenario}

In our research, based on our discussions with dissident journalists, we considered a practical scenario where the victim is a high-profile target who relies on secure, popular, end-to-end encrypted messaging applications (i.e., WhatsApp and Signal) for official and private communication using their smartphone.  
We assume that the attacker knows the victim's phone number and is equipped with a powerful zero-click exploit code that leverages a zero-day vulnerability in one of the installed chat applications (e.g., WhatsApp) to compromise the smartphone without the victim's interaction. 
The attacker aims to attain unfettered access to the victim's smartphone and covertly extract information such as chats, contacts, keystrokes, and location or record real-time audio and video. 
The above threat mirrors recent revelations regarding the secret surveillance of high-profile targets using the Pegasus spyware.

One question that may come up is why the victim needs to use WhatsApp and cannot simply resort to secure email? 
This is because WhatsApp is ubiquitous, provides good security against eavesdropping by authorities, and in the case of dissident journalists, is frequently already used by their sources.

\subsection{Desirable Characteristics}
First, we identified the necessary properties that would make a system zero-click resistant (i.e., even in the case of a successful zero-click exploit, the information leaked to the adversary is significantly limited). 
We envision these characteristics as essential to provide ideal security against zero-click attacks with minimal system and attacker assumptions. 

\begin{itemize}
%[leftmargin=*]
\item \textbf{No direct delivery of messages to the smartphone:} 
Zero-click exploits typically abuse zero-day vulnerabilities in chat applications to compromise a device.
The mere delivery of a message containing the payload exploit is sufficient to compromise the victim's smartphone.  
Therefore, it is necessary to prevent any message from being delivered directly to the smartphone. 

\item \textbf{Remote access to messages:} 
Since the messages are not delivered directly to the smartphone, one idea would be to provide the user with a mechanism to interact with the messages remotely.  
It is important to note that the messages cannot be delivered to the device at any point, not even temporarily. 
This is in contrast to using a web application to access these messages. 
For example, opening an image sent over email in the web interface would still "download" the image on the device. 
Zero-click exploits can potentially evade input sanitization, so any malicious zero-click code can escape the browser's sandbox and access the device's data. Therefore messages need to be accessed completely in place outside the victim device. 

\item \textbf{Temporal and spatial application isolation:} 
We believe that the solution should provide temporal and spatial isolation between applications. 
This is because zero-click exploits targeting a specific chat application (e.g., WhatsApp) can potentially abuse cross-application vulnerabilities to infect other installed applications (e.g., Signal). 
Hence, it would be necessary to isolate individual chat applications to limit the impact of the attack to the vulnerable application only.
Moreover, the framework would need to limit the lateral damage of a successful zero-click attack and prevent the attacker to passively eavesdrop the vulnerable chat application for a long time.

\item \textbf{Scalable and usable:} 
Finally, it is likely that the user needs to add more third-party chat applications to the remote server in the future.
%an organization will have several high-risk individuals who need protection. 
Hence, the system must be scalable with respect to the increase in the number of applications.  
For complete adoption, usability plays a critical role. 
Therefore, the system would need to facilitate seamless access to messages over the Internet without requiring significant user interaction. 
\end{itemize}

\begin{figure}[t]
\begin{center}
\includegraphics[width=\linewidth]{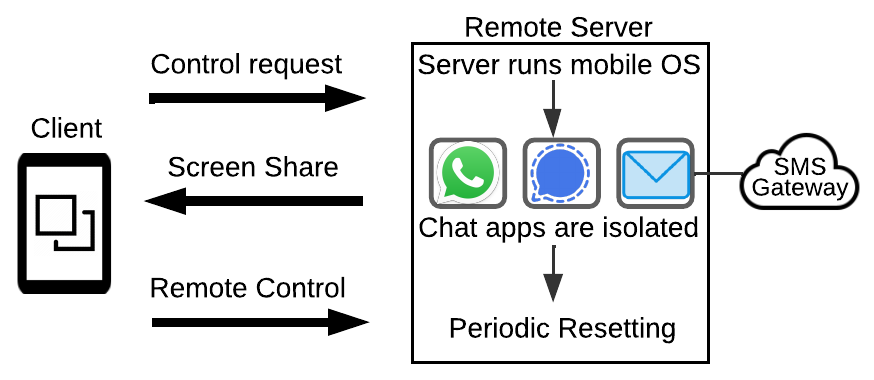}
\caption{Experimental Zero-Click Secure Architecture: User accesses remote isolated applications via screen sharing and remote control mechanism. Regular messages are routed through cloud SMS Gateway, and the server is periodically reset to the initial unaffected snapshot to remove any zero-click infection (if present).}
\label{fig:design}
\end{center}
\end{figure}

\subsection{An Experimental Zero-click Secure Architecture}
Based on the above properties, we attempted to design a zero-click secure architecture that we present below.  
One of the primary motivations for the design was to be able to realize it with off-the-shelf components. 
Since a single zero-click exploit is potent enough to gain complete access to the smartphone without any user interaction, it is imperative to block (or reduce) all possible attack vectors.
Unfortunately, zero-click exploits leverage unpatched vulnerabilities, making it difficult for smartphone users or even anti-malware solutions to detect or prevent such attacks.  
The core idea of our attempted secure framework (as illustrated in Figure \ref{fig:design}) was to shift the attack surface from the user's device (hereby called client) to a virtual smartphone (such as cloud or remote server) that runs all the sensitive applications.

Since chat applications, by default, parse messages even from unknown/ untrusted numbers, they are an obvious zero-click attack vector. 
Hence, all chat applications must be shifted to the remote server such that the user has unattended access to the remote applications at all times. 
Third-party chat applications such as WhatsApp and Signal can be configured on the server effortlessly, as they only require one-time authentication using a one-time password while registering the application.
However, shifting in-built messaging applications from the client device to the remote server is challenging.  
We leveraged a cloud-based SMS Gateway, which uses HTTPS messages to enable any device (even computers) to send and receive SMS over the telecommunication network. 
To recall, shifting chat applications remotely prevents the zero-click exploits from reaching the client device and activating its hardware for real-time surveillance.

It is pertinent to note that moving the chat applications to a remote server alone is insufficient. 
Zero-click exploits can evade message sanitization; hence, even if the server sanitizes each received message before forwarding it to the client device, there is an obvious risk of the zero-click exploit being forwarded to the device. 
Our secure framework allowed the user to indirectly access the chat applications on the remote server using a user-intuitive screen sharing and remote control mechanism. 
Unlike sending sanitized input to the client, screen sharing takes screen pixels and shares them with the client, thereby removing any possibility of the malware reaching the client device.  
In addition, the remote control feature allows the user to control the chat applications remotely.  
If the user wishes to send a message, the user's input (keystrokes) on the mirrored screen is forwarded to the remote server and then to the intended recipient.

Our design rationale was that screen sharing considerably reduces the window of attack opportunity and protects the client device against zero-click attacks; however, as a zero-click exploit targeting a specific application can also exploit cross-application vulnerabilities, the \emph{remaining applications on the server} are also susceptible to the zero-click attack. 
To limit the attack's impact, we run each application in isolation in a sandbox-like environment, for instance, a separate virtual machine or docker container.
So far, there is no evidence of \emph{cross-OS exploitation} for zero-click exploits, as the existing zero-click exploits are specifically designed to abuse a certain zero-day bug in a smart application installed on the mobile OS. 
Hence, for example, the zero-click exploit cannot exfiltrate from the Android instance A running vulnerable application A to the non-mobile host OS (e.g., Linux server) and into another Android instance B to compromise application B.  
With this setup, the zero-click attack, if launched, should be restricted to the targeted smart application only (i.e., application A).

Furthermore, to limit the time of successful zero-click exploitation and prevent the attacker from passively eavesdropping on the targeted chat application for a long time, we thought about periodically resetting each instance to the initial unaffected snapshot (e.g., every three days or as required by the user). 
Our key intuition here was that this would help terminate any malicious connection established by the attacker, as forensic investigations of Pegasus-infected smartphones have proved that Pegasus avoids persistence to evade anti-malware detection, and resetting the smartphone to factory settings removes the spyware~\cite{persistence}. 
Hence, this would require additional effort from the attacker to re-infect the target application and would increase the difficulty bar.

In essence, the fundamental idea of our secure framework was to subside the risks and damages caused by a zero-click infection by shifting the attack surface from the client to the remote server and allowing the user to access remote, isolated chat applications via screen mirroring only. 
\section{Framework Realization}
To determine the feasibility of implementing the proposed framework using COTS components, we tried and tested several approaches to run the mobile OS remotely, isolate applications, and interact with the remote server.
We came across several challenges and learned a number of lessons while building this secure framework that we discuss in Section~\ref{lessons}. 
Below, we walk through the approach that we found to be the most feasible that we then implemented on an Android-based remote server.
We specifically chose Android because of its open-source and more accessible nature. 
Irrespective of the server OS, conceptually speaking, the client OS can be Android or iOS.
We assumed that the computing resources required to realize the secure framework (i.e., client device, remote server, screen sharing protocol and SMS Gateway) were not vulnerable to any other cyber or physical attack, and the only risk in question was zero-click exploitation.

\begin{figure}[t]
\begin{center}
\includegraphics[width=\linewidth]{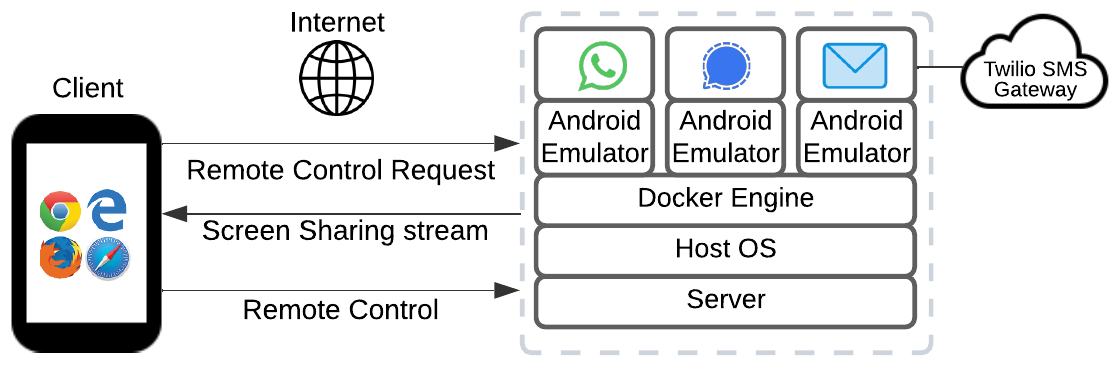}
\caption{Framework Realization: We ran each chat app (WhatsApp, Signal and messaging application) remotely in containerized Android Emulators on GCP and accessed them from the mobile's web browser via WebRTC service.}
\label{fig:solution2}
\end{center}
\end{figure}

\subsection{Server Setup and Application Isolation}
We set up an Android cloud emulator~\cite{androidcloud} on Google Cloud Platform (GCP) that enabled us to run the official Android Emulator~\cite{androidstudio} on top of Docker~\cite{docker} as a web service.  
Android Emulator is part of Google's development environment for Android, i.e., Android Studio, which enabled us to run Android instance on the cloud. 
In addition, Docker is an open-source platform that uses OS-level virtualization to package application(s) in a container, i.e., isolate applications from the host system.
We configured each chat application (WhatsApp, Signal and regular messaging application) to run in a separate Docker container (as illustrated in Figure~\ref{fig:solution2}), which considerably reduced the chances of cross-application zero-click exploitation. 
WhatsApp and Signal were configured on the server using the user's number.
To receive SMS/ regular messages on the remote server, we configured the messaging application with the cloud-based SMS Gateway service, namely Twilio~\cite{twilio}.  
Twilio API forwarded the SMS to and from our server over HTTPS (TLS 1.2 encryption)~\cite{twiliosecurity}, without the need of a physical SIM card. 
The API can be further tweaked to allow messages only from verified contacts and disable the preview of web links in the text messages, thereby reducing the chances of zero-click infection.
This setup has an added advantage; iMessages are received as regular SMS on the server, and the client device (i.e., iPhone) remains protected from any iMessage-based exploit.

As Android Emulator requires Kernel-based Virtual Machine (KVM) to run Android OS smoothly, the hardware-based virtualization enabled us to implement this setup on our dedicated Linux server as well as cloud platforms, i.e., GCP, Amazon Web Services (AWS) and Microsoft Azure.  
In short, containerized Android Emulator helped run chat applications remotely in isolation.

\subsection{Remote Interaction with Chat Applications} 
The Android Cloud Emulator~\cite{androidcloud} utilized WebRTC protocol~\cite{webrtc} to display the remote screen on the client device.  
WebRTC is an open-source project that provides \textit{real-time communication} capabilities by supporting video, voice, and data for the web. 
As WebRTC does not require installation of any plugin or third-party software at either end, it essentially helped obtain unattended access to the server, i.e., it bypassed the need for a human agent to accept the connection request at the server every time and ensured connectivity even after a system restart.
However, WebRTC requires a direct connection between the communicating peers, which is not always possible over the Internet.  
Hence, we utilized a TURN (i.e., Traversal Using Relays around NAT) server to relay the contents of the remote screen to the client device over the Internet. 
This made the server screen accessible via the server's public IP. 
Note that iOS does not fully support screen sharing via WebRTC on the mobile's web browser; hence, we configured screen sharing via an alternate method that exchanges png screenshots to provide screen capturing for iOS users.

Next, for remotely controlling the server screen from the client device, we used a React web application that helped deliver the user's keystrokes on the mirrored screen (on the client device) to the server.
To improve the user experience, we modified the React web application to provide a full-screen view. 

\subsection{Other Security Features} 
To proactively limit the damage caused by a successful zero-click attack and prevent the attacker from passively eavesdropping on the targeted application for a long time, we decided to periodically reset each instance to its initial (unaffected) snapshot. 
Our thinking was that this would help wipe off undetected infections and terminate connections (if any) established by the attacker. 

Moreover, as each remote instance serves as an independent Android smartphone, we enabled Kiosk mode to restrict each container to the predefined application so it can remain open at the server 24/7.  
GoKiosk~\cite{gokiosk} is one of the Kiosk applications that is freely available on the Play Store.  
In short, with this setup, the client device did not receive any untrusted message directly; rather, the message was first delivered to the server and then streamed via screen mirroring on the client device.
Hence, our key insight was that the zero-click attack could only compromise the targeted application and not affect the client device or other remote apps.
\section{Performance and Usability}
~\label{testing} 
To gain insights into our design and distill some lessons, we evaluated the performance and usability of our off-the-shelf, experimental framework in two ways; i) System Testing (frequently carried out by our team), and ii) Quality Assurance Testing (carried out by prospective users).  
Note that the framework's security analysis requires sending a zero-click exploit to the victim and forensically investigating if the victim's device has been infected or not.
We could not test the framework from a security perspective owing to the absence of zero-click binaries. 
However, the fact that the chat applications were sandboxed/isolated means the design principle of separation of privilege secures them.  

\begin{table}[t]
\centering
\caption{Evaluating connection establishment phase with different client OS, server deployments, and networks.}
\label{tab:conntime}
\resizebox{\columnwidth}{!}{%
\begin{tabular}{l|l|l|c|c|c|c|c} 
\hline
\multicolumn{1}{c|}{\multirow{2}{*}{\begin{tabular}[c]{@{}c@{}}\textbf{Client }\\\textbf{OS}\end{tabular}}} & \multicolumn{1}{c|}{\multirow{2}{*}{\begin{tabular}[c]{@{}c@{}}\textbf{Access}\\\textbf{Approach}\end{tabular}}} & \multicolumn{1}{c|}{\multirow{2}{*}{\begin{tabular}[c]{@{}c@{}}\textbf{Connection}\\\textbf{Status}\end{tabular}}} & \multicolumn{3}{c|}{\textbf{Wi-Fi}}  & \multicolumn{2}{c}{\textbf{Cellular}}  \\ 
\cline{4-8}
\multicolumn{1}{c|}{}                                                                                       & \multicolumn{1}{c|}{}                                                                                            & \multicolumn{1}{c|}{}                                                                                              & \textbf{A} & \textbf{B} & \textbf{C} & \textbf{A} & \textbf{B}                \\ 
%\hline
%\multicolumn{8}{c}{\textbf{GCP Cloud Server}}                                                                                                                                                                                                                                                                                                                                                                                       \\ 
\hline
\multirow{4}{*}{Android}                                                                                & \multirow{4}{*}{\begin{tabular}[c]{@{}l@{}}WebRTC\end{tabular}}                                          & Successful                                                                                                         & 25         & 25         & 25         & 25         & 25                        \\ 
\cline{3-8}
                                                                                                            &                                                                                                                  & Failed                                                                                                             & 0          & 0          & 0          & 0          & 0                         \\ 
\cline{3-8}
                                                                              &                                                                                                                  & Re-tried                                                                                                           & 0          & 0          & 0          & 0        & 0                         \\ 
\cline{3-8}
                                                                                              &                                                                                                                  & Terminated                                                                                                         & 0          & 0          & 0          & 0          & 0                         \\ 
\hline
\multirow{4}{*}{iOS}                                                                                        & \multirow{4}{*}{\begin{tabular}[c]{@{}l@{}}Screen\\Capturing\end{tabular}}                                           & Successful                                                                                                         & 25         & 25         & 25        & 25         & 25                        \\ 
\cline{3-8}
                                                                              &                                                                                                                  & Failed                                                                                                            & 0          & 0          & 0          & 0          & 0                         \\ 
\cline{3-8}
                                                                                  &                                                                                                                  & Re-tried                                                                                                           & 0          & 0          & 0          & 0          & 0                         \\ 
\cline{3-8}
                                                                                                  &                                                                                                                  & Terminated                                                                                                         & 0          & 0          & 0          & 0          & 0                         \\ 
\hline
\end{tabular}}
\end{table} 
\subsection{Setup} We performed tests on eight different smartphones to ensure that our implementation was independent of the screen sizes, device manufacturers and web browsers. 
The Android phones included LG V40 ThinQ, Motorolla Nexus 6, Samsung Galaxy S10E, Oppo A5 and Vivo U1, whereas the iOS devices included iPhone XR, iPhone X and iPhone 7. 
In addition, we also tested other device models by creating virtual devices in the Android Emulator~\cite{androidstudio}.  
We set up the server side on a Linux-based VM instance on GCP. 
This instance had nested virtualization enabled and was accessible over the Internet using public IP. 
We ran three Docker containers on GCP, one for each application (WhatsApp, Signal and regular messaging application) and configured the applications using Google Voice number.  
For testing, we set up three more GCP accounts and configured the applications with different Google voice numbers. 
This essentially allowed the participants of the user study to test the framework without linking their personal accounts on the server before they were fully satisfied. 
To encourage active involvement from the participants, group chats were also created where random texts, images, gifs, etc., were frequently shared.

\subsection{System Testing} 
\label{performance} 
Here, we assessed our system's connection time, usability, introduced latency, and scalability. 

\paragraph{Connection Time:} 
To evaluate the connection establishment phase, we accessed the remote server on several phones at different times, using various networks. 
For accessing the remote server, we specifically used the WebRTC-based screen sharing approach for Android clients and the PNG-based screen sharing approach for iOS clients.  
We made 25 connection attempts from the client device to the remote server and observed how often the connection fails, how many attempts are required to reconnect, and if the connection ever terminates itself while the system is being used. 
Each round of the experiment lasted for an hour. Our results (Table~\ref{tab:conntime}) indicate that the client was able to connect to the server in the first attempt seamlessly, and the connection was stable throughout the experiment. 

\paragraph{Usability:} 
To evaluate our experimental system's usability, we accessed the remote server from smartphones of different screen sizes, device
manufacturers, and web browsers (notably Google Chrome, Firefox, Microsoft Edge, Opera, UC and Oppo browser). In all instances, the text was readable, the remote display was of high quality and covered maximum screen size. 
However, as our framework added an additional layer to the communication path, it introduced a lag which sometimes created usability challenges. 

\paragraph{Measuring Lag:}
In practice, it is difficult to measure the introduced lag precisely, as the clocks of the two computing devices (phone and server) are not in sync. 
Therefore, we measured the additional transmission time between client and server as half of the Round Trip Time (RTT) between these two nodes. 
For this, we developed client and server Android applications, whereby the client sends a test message to the server, which replies with an acknowledgement (ACK).
The client then measures RTT as the difference between the time the message was sent, and the ACK was received. 
To minimize the measurement error, the client sends ten messages per second and calculates the average lag. 
Although the RTT measurement is influenced by other factors such as network speed, traffic load, etc., the measurements provided a quantitative value to the introduced lag.
Table~\ref{tab:delay} reports the average lag in seconds (s) over different networks. 
One important question here was, given the security benefits of our solution and that the user experience was not always significantly affected, whether the 0.49 seconds lag is acceptable?  \begin{table}
\centering
\caption{Measuring lag introduced by COTS components at different times and network connections.}
\label{tab:delay}
\begin{tabular}{l|c|c|c|c|c} 
\hline
\multicolumn{1}{c|}{\multirow{3}{*}{\textbf{Client Device}}} & \multicolumn{5}{c}{\textbf{Introduced Lag (s)}}                                                                                           \\ 
\cline{2-6}
\multicolumn{1}{c|}{}                                        & \multicolumn{3}{c|}{\textbf{Wi-Fi}}                                               & \multicolumn{2}{l}{\textbf{Cellular}}                 \\ 
\cline{2-6}
\multicolumn{1}{c|}{}                                        & \textbf{A}                & \textbf{B}                & \textbf{C}                & \textbf{A}                & \textbf{B}                \\ 
\hline
LG V40 ThinQ                                                 & 0.44                      & 0.47                      & 0.42                      & 0.49                      & 0.51                      \\ 
\hline
Motorolla Nexus 6                                            & 0.45                      & 0.51                      & 0.43                      & 0.50                      & 0.53                      \\ 
\hline
Samsung Galaxy S10E                                          & 0.47                      & 0.52                      & 0.49                      & 0.53                      & 0.56                      \\ 
\hline
Oppo A5                                                      & 0.42                      & 0.53                      & 0.54                      & 0.54                      & 0.55                      \\ 
\hline
Vivo U1                                                      & \multicolumn{1}{l|}{0.45} & \multicolumn{1}{l|}{0.48} & \multicolumn{1}{l|}{0.49} & \multicolumn{1}{l|}{0.48} & \multicolumn{1}{l}{0.50}  \\ 
\hline\hline
\textbf{Average Lag (s)        }                                      & \multicolumn{5}{c}{ 0.49}                                                                                                                 \\
\hline
\end{tabular}
\end{table} 

\begin{figure}[t]
\begin{center}
\includegraphics[width=\linewidth]{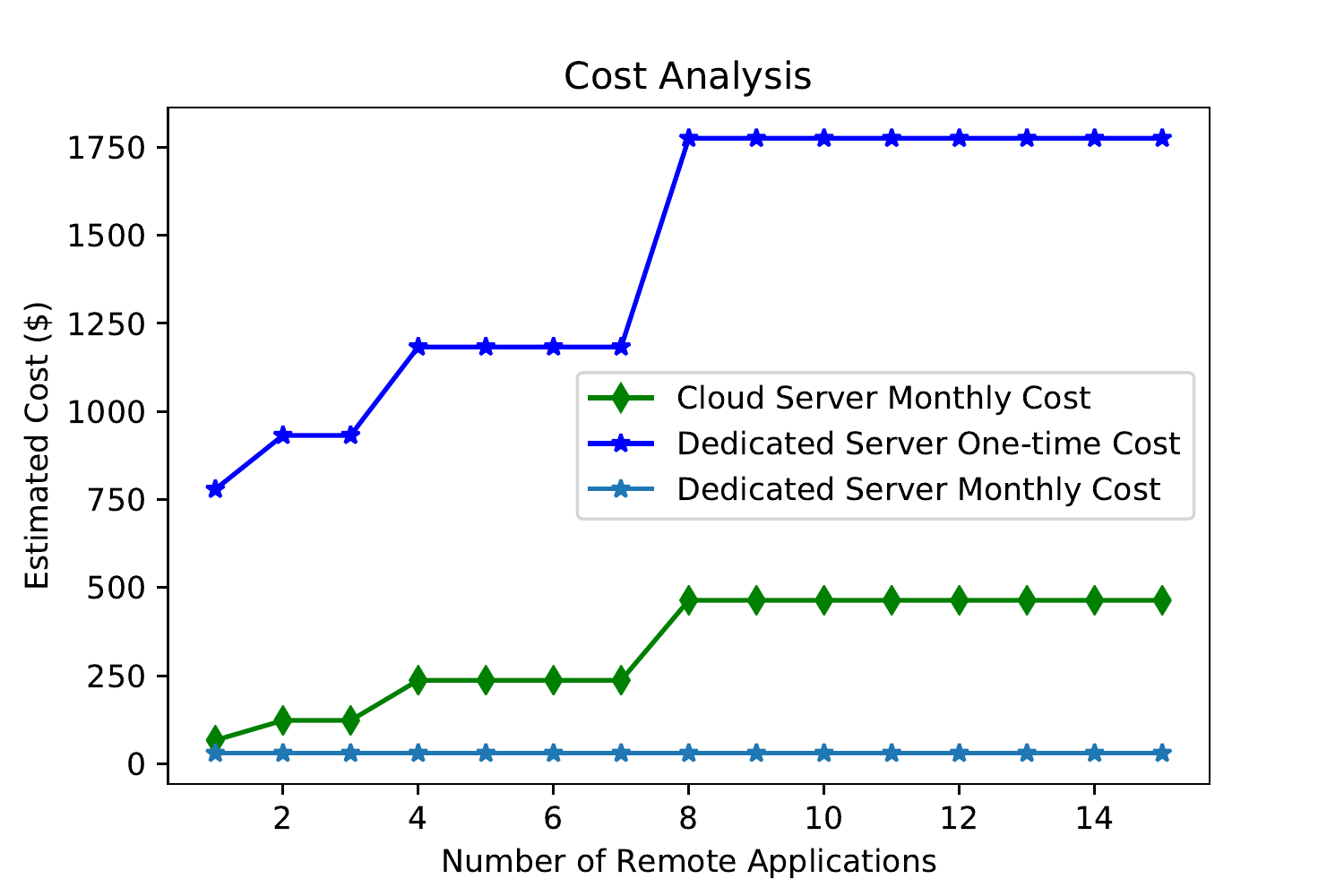}
\caption{Cost vs resource analysis per user for deploying server end on Google Cloud Platform and dedicated server.}
\label{fig:plot}
\end{center}
\end{figure}

\paragraph{Resource Requirement and Scalability:} 
Realistically, the user might wish to port additional chat applications (e.g., Viber, Telegram) remotely. 
To determine if our COTs-based framework was scalable, we performed cost analysis for a single user with respect to the increase in the number of remote applications.  
We observed that for efficient performance, the bare-minimum requirement for each Android emulator to run a chat application is 4GB RAM and 2GB disk space.  
Hence, we kept the GCP cloud instance with the Intel Haswell CPU Platform, 8GB RAM and 100GB disk space as the baseline for running one chat application.
As disk space is not a hard constraint, we noted the monthly cost as we increased the required RAM (16, 32 and 64 GB) corresponding to the number of applications to be hosted.  
Note that the user can also self-deploy the server side to lessen the recurring expenditure. 
To provide cost analysis for a dedicated server, we kept a Dell PowerEdge server with 8GB RAM and 1TB disk space (\$780) as the base server for running one chat application and kept increasing the RAM as the number of applications increased. 
The results in Figure~\ref{fig:plot} suggest that despite high initial investment, the dedicated server is more cost-effective than a cloud setup in the long run.
However, our rationale was that the GCP cloud setup is practically more secure in terms of the server's physical and logical security.

\subsection{Quality Assurance Testing} \label{usability}
To determine our envisioned system's usability in practice, we got the system evaluated by potential users and received feedback.

\paragraph{Ethical Considerations:} 
We note that our usability study was exempted under the IRB Exemption Category 3 - Benign Behavioral Interventions by our institution's IRB.  
Although our study involves human subjects, it does not collect sensitive and personal information or perform deception, attacks, etc.

To recruit participants for the user study, we sent out invitation emails to 30 individuals from our personal and professional circle belonging to different regions of the world.  
On a high level, all participants were above 18 years and included i) potential targets of zero-click attacks (including three journalists) and ii) privacy-conscious smartphone users.  
Of 30, 27 individuals showed willingness to participate in the user study, whereas 3 individuals did not respond.
As the user study was voluntary, we did not send follow-up emails. 
We scheduled Zoom meetings with the participants to obtain informed consent, demonstrate the working of our system and provide instructions regarding the user study.  
Since participants might initially have privacy concerns regarding setting up their personal chat accounts on the remote server, we pre-configured the applications on each GCP test account using Google voice numbers.
We did not collect personally identifiable information (e.g., name or email of the participants) or other related data like IP addresses during the study.
To maintain the confidentiality of the results, we analyzed and reported the responses in our work as group data without identifying any individual.
Finally, after the user study, we deleted all the conversations from the server. 

\paragraph{User Study and Results:}
The tasks involved accessing the remote server by typing the provided public IP of the server in the mobile web browser.
The participants were given temporary login credentials to prevent unauthorized access to the remote server. 
Once logged in, each participant was asked to exchange random text messages, images, gifs, etc., with the saved contacts or engage in group chats.  
Our team controlled the other contacts and group chats to let the participant actively communicate using the system and observe if they experienced any possible delay or performance degradation.
After using the system for an hour, participants shared their feedback on the following via a brief Google survey form (provided in the Appendix).  

\begin{enumerate}

\item Prior know-how of zero-click attacks, 

\item Experience connecting to the remote server, i.e., Is it accessible? Is the connection stable or terminates frequently? 

\item Experience of sending test messages, i.e., Is the system user-friendly? Is the lag acceptable? 

\item Suggestions to improve the system.

\end{enumerate}

\begin{figure}[t] \begin{center}
\includegraphics[width=\linewidth]{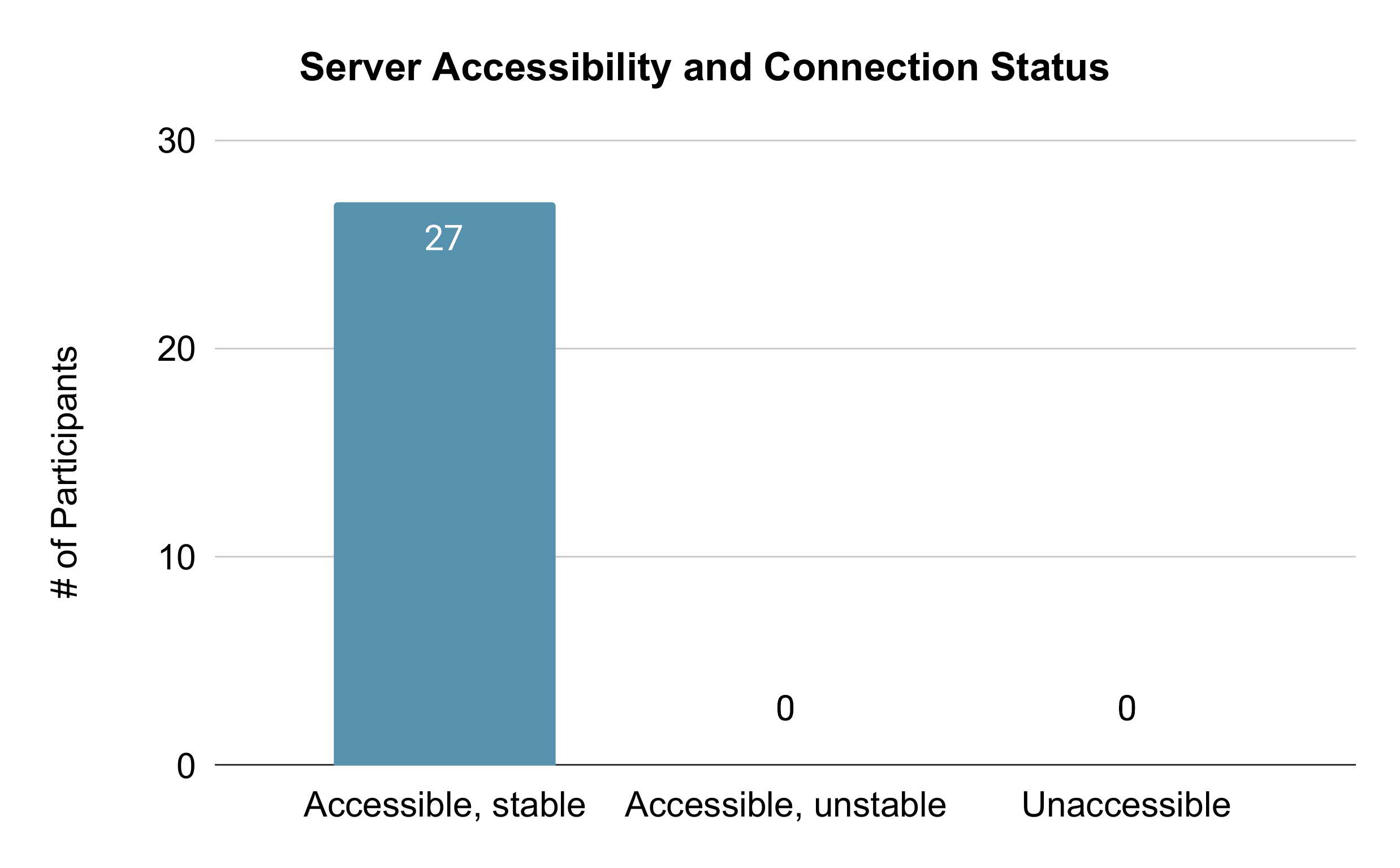} \caption{Evaluating Server's Accessibility and Connection Status via User Study: All 27 participants found the server accessible in the first attempt and the connection stable.}
\label{fig:conn} \end{center} \end{figure}

The results indicated that 21 out of 27 participants had a fair idea about zero-click attacks before starting the user study.  
Figure~\ref{fig:conn} shows the results of the participant's experience while connecting to the remote server.
As shown, for all 27 participants accessing the server from different regions at different timings, the remote server was accessible in the first attempt, and the connection remained stable throughout the study (i.e., it did not terminate at all).
  
\begin{figure}[t]
\begin{center}
\includegraphics[width=\linewidth]{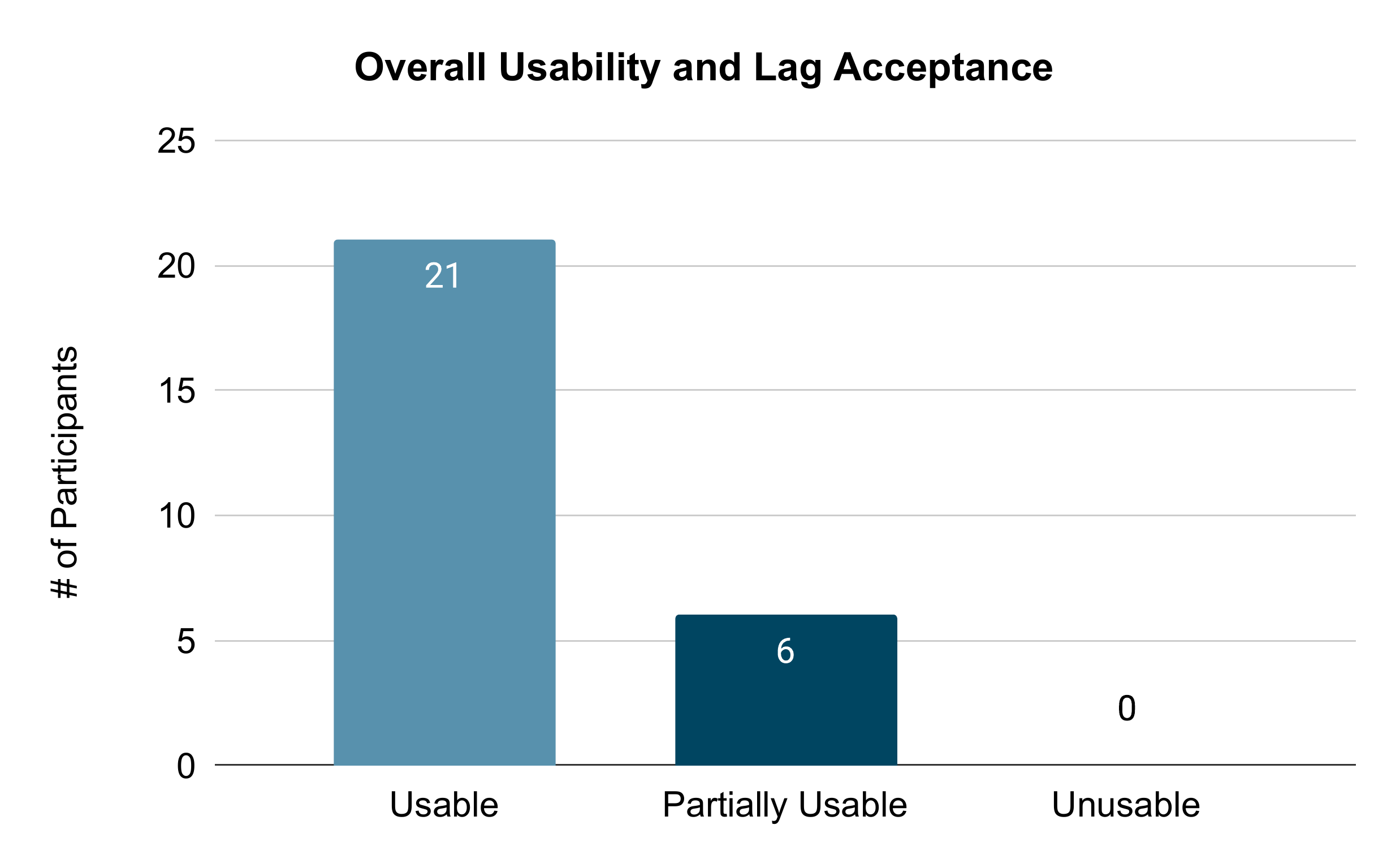}
\caption{Evaluating System's Usability via User Study: Here, usable = user-friendly and acceptable lag, partially usable = user-friendly but unacceptable lag, unusable = not friendly and unacceptable lag. Result: 21 participants found the system usable, while 6 participants deemed it partially usable.}
\label{fig:usability}
\end{center}
\end{figure}

Figure~\ref{fig:usability} indicates the results of the participant's experience while sending test messages. 
For this, we specifically evaluated if the participants found the system user-friendly (i.e., text was readable, the screen was full size, etc.) and whether the lag introduced by the COTS-based components was acceptable considering the security benefits of the solution. 
The results indicated that 21 out of 27 participants found the system to be user-friendly and the lag (interestingly) acceptable.
Six participants found the system to be user-friendly but complained that the lag was significant. 
Overall, the participants gave constructive feedback, such as keeping the native keyboard on-screen while controlling the remote screen and ensuring that the user does not experience any lag.

\section{Experiences and Lessons Learned} \label{lessons}

In this section, we summarize the lessons we distilled from our experience of designing and implementing a secure, off-the-shelf and readily-available framework against zero-click attacks. 
We hope that these lessons and the insights from real users will be useful for researchers and mobile developers who wish to develop more rigorous mobile security solutions.

\subsection{Running Mobile OS on the Server} 

\subsubsection*{{\textbf{Lesson Learned \# 1:}}} 
\textit{{There are only a few open solutions to virtually run the mobile OS 24/7 in portrait orientation, and the most usable and scalable option appears to be the official Android Emulator.}} 

The first and foremost task in setting up a virtual smartphone was to run the mobile OS on our Linux-based dedicated server.  
The basic requirement was to find an appropriate option that is free, reliable, updated, and provides display in portrait mode. 
To this end, we explored existing local and cloud-based emulators, simulators, and virtual environments for Android and iOS.

Since Apple only allows iOS to run on iOS devices and the Xcode simulator~\cite{xcode}, it was difficult to emulate the iOS environment at the remote end where the underlying OS was Linux.
Likewise, existing web-based iOS emulators, namely Appetize.io~\cite{appetize} and Corellium~\cite{corellium}, also use an iOS device at the backend to feature a virtual iOS device and are costly.
Such options are only feasible for \textit{testing} iOS apps and are not meant for providing 24/7 service.

In contrast, there are several public and proprietary solutions that emulate an Android device. 
Below we discuss our experience using these options in the light of our requirements, followed by its summary in Table~\ref{tab:mobileos}.  
The official Android Emulator~\cite{androidstudio} provided a stable environment to run Android OS.  
Unfortunately, Android Emulator had high system requirements and limited mobile devices that support Google Play Store. 
In addition, the mobile versions with Play Store were restricted to automatic graphics, whereas, for other devices, we could choose hardware or software graphics, vary heap size, etc.
This adversely affected the performance of Android applications in Play Store-compatible emulated devices.  
However, as we ran only one application per Android emulator, these issues did not compromise the user experience.

We also explored the option of freely available, unofficial Android gaming emulators such as Memu~\cite{memu} and BlueStacks~\cite{bluestacks}.  
These emulators were unreliable, had sluggish performance, and were often stuck on the boot.  
BlueStacks and Memu displayed many advertisements and were often detected to be malicious by antivirus solutions; hence, we disregarded these unreliable emulators. \begin{table}[t]
\centering
\caption{Existing solutions to run mobile OS on a remote instance and their limitations}
\label{tab:mobileos}
%\resizebox{\columnwidth}{!}{%
\begin{tabular}{l|l|l} 
\hline
\#                                          & \multicolumn{1}{c|}{\textbf{Solution}}                                                         & \multicolumn{1}{c}{\textbf{Limitations}}                                                                                                \\ 
\hline
1                                           & Xcode simulator~\cite{xcode}& Requires iOS device, costly,                                   \\ 
\hline
2                                           & \begin{tabular}[c]{@{}l@{}}iOS web emulators \\ (Appetize~\cite{appetize} and \\Corellium~\cite{corellium})\end{tabular}                              & \begin{tabular}[c]{@{}l@{}} Runs for limited duration, \\Costly,       \end{tabular}                             \\ 
\hline
3                                           & \begin{tabular}[c]{@{}l@{}}Official Android \\ Emulator~\cite{androidstudio}\end{tabular}                              & \begin{tabular}[c]{@{}l@{}}High system requirements, \\Limited devices with PlayStore,\\\end{tabular}                                   \\ 
\hline
\begin{tabular}[c]{@{}l@{}}\\4\end{tabular} & \begin{tabular}[c]{@{}l@{}}Android Gaming \\ Platforms (Memu~\cite{memu}, \\BlueStacks~\cite{bluestacks}, etc.)\end{tabular} & \begin{tabular}[c]{@{}l@{}}RAM hungry, sluggish, \\Stucks on booting, \\Often detected as virus, \\Too many advertisements,\end{tabular}  \\ 
\hline
5                                           & Anbox~\cite{anbox}                                                                                          & \begin{tabular}[c]{@{}l@{}}No portrait mode, \\No built-in app store, \\Runs on a few Linux distros,\end{tabular}                         \\ 
\hline
6                                           & GenyMotion~\cite{genymotion}                                                                                    & \begin{tabular}[c]{@{}l@{}}Costly, \\No built-in app store, \\Not GApps compatible,\end{tabular}                                     \\ 
\hline
7                                           & Android-x86 VM~\cite{androidx86}                                                                                 & \begin{tabular}[c]{@{}l@{}}Doesn't boot on cloud, \\Desktop mode by default, \\Resource intensive, \end{tabular}                                               \\ 
\hline
8                                            & Cuttlefish~\cite{cuttlefish}             & \begin{tabular}[c]{@{}l@{}}Very resource intensive,\\Specific system requirements,\end{tabular}                                 \\
\hline
9                                           & \begin{tabular}[c]{@{}l@{}}Windows 11 \\ Android Subsystem~\cite{androidwindows}\end{tabular}                        & \begin{tabular}[c]{@{}l@{}}Not stable, frequent flashbacks,\\Specific system requirements.\end{tabular}                                 \\
\hline
\end{tabular}
\end{table}

Besides, we also tested common Android emulators, Anbox~\cite{anbox} and GenyMotion~\cite{genymotion}, that can run Android applications in the cloud. 
Anbox is a stable option, but since it is a desktop emulator, it cannot run in portrait mode, making it difficult to view the remote landscape screen on the smartphone even with screen rotation enabled.  
In contrast, GenyMotion uses a custom AOSP ROM, and takes advantage of OpenGL-capable graphics cards to run Android OS on the cloud platform in portrait orientation smoothly. 
However, Genymotion was expensive (i.e., \$0.5 per hour per instance) and hence, not a practical option for a server that has to run multiple instances 24/7. 
Also, GenyMotion did not have a built-in Google Play Store, and was incompatible with most Google Apps (GApps). 

Android-x86 OS~\cite{androidx86}, a port of the Android Open Source Project (AOSP), is another powerful option that can run Android OS on Intel x86 or AMD-powered devices.  
Android-x86 can be installed on a physical or virtual machine.  
We installed it in VirtualBox~\cite{virtualbox} to virtualize the Android instance on our Linux-based dedicated server, with KVM enabled.  
By default, Android-x86 offers landscape orientation. 
We modified the GRUB boot loader to change the orientation to portrait mode, making it feasible to fit the remote screen on the user's smartphone.  
However, as VirtualBox runs a full copy of OS and a virtual copy of requisite hardware, the setup was resource-intensive for merely running a single chat application.

In contrast to the official Android Emulator, Android's Cuttlefish~\cite{cuttlefish} is a configurable virtual Android device that responds to the application interactions at the OS level, just like a physical phone.
Cuttlefish can run locally on Linux x86 computing machines and cloud offerings such as GCP. 
However, Cuttlefish supports API levels after 28, runs only on a few Debian-based distributions, and is extremely resource-intensive.  
For us, the installation alone took more than 8 hours and a minimum of 250 GB disk space, making it an infeasible option for setting up multiple Android instances.

Lastly, we tested Microsoft's Windows 11 subsystem for Android~\cite{androidwindows} which makes it possible to run Android applications on Windows 11. 
This is a promising solution; however, the Android applications continued to crash while running in the beta phase.  
To conclude, the official Android emulator seemed to be the most appropriate choice to run mobile OS at the server end reliably. 

\subsection{Sandboxing Applications on the Server}

\subsubsection*{{\textbf{Lesson Learned \# 2:}}} \textit{{Android's default
Application Sandbox is ineffective against zero-click cross-application attacks.
Hence, virtualization or containerization techniques can be employed to isolate applications with some security and resource trade-offs.}} 

Android OS has a default Application Sandbox~\cite{sandbox} that works by assigning a unique User ID (UID) to every Android application and runs each application in isolation as a separate process. 
This kernel-level sandbox extends protection to OS applications, native code and everything above the kernel, including OS libraries and application framework. 
Despite this, zero-click exploits can defeat the Android Application Sandbox and escalate privileges to gain root access to the entire device. 
Hence, we investigated more powerful isolation mechanisms, i.e., virtualization and containerization, to isolate Android applications on the server safely.

We ran individual applications in Android Emulator on top of separate virtual machines and believe that it provides ultimate security against zero-click-based cross-application exploits and cross-OS exploits.  
This is because a zero-click exploit cannot exfiltrate from Android instance A running application A to the Linux host and into Android instance B to compromise application B. 
Although secure, this setup was extremely resource-intensive.  
We also tested multiple instances of Android Emulator on top of Docker~\cite{docker}, which essentially allowed us to isolate each application in a separate container. 
Like virtualization, containerization also reduced the prospective susceptibility of the installed applications to invisible zero-click attacks; however, this setup was much more lightweight, and hence, was preferred.

\subsection{Implementing the Server on Cloud Platform}

\subsubsection*{{\textbf{Lesson Learned \# 3:}}} \textit{{Mobile OS requires hardware virtualization for enhanced performance and graphics; hence, running the mobile OS on bare-metal cloud servers is expensive and not a long-term and scalable solution. 
An appropriate trade-off is to employ software acceleration to run the mobile OS on public cloud platforms.}}

After successfully implementing containerized Android Emulator setup on our dedicated Linux server, we tested its feasibility in the cloud platforms, namely GCP, AWS, and Azure. 
Although the Android Emulator was successfully installed in the cloud, it showed sluggish performance. 
In general, the emulator's performance can be improved by enabling virtual machine and graphic acceleration~\cite{virtualization} forcing the hypervisor to use the processor of the underlying computer to improve the emulator's execution speed. 
As buying a bare-metal cloud instance or adding a GPU are not cost-effective options to run a mobile OS 24/7, we instead relied on enhancing the emulator's performance through software acceleration (where the Android Emulator simulated GPU processing using the computer's CPU).
This enabled us to run Android Emulator in Docker containers on the cloud with acceptable performance. 

Note that we also tested the feasibility of another viable alternative, i.e., Android-x86 OS-based VM on cloud platforms. However, it was unstable (i.e., it could not boot up or kept crashing) even with nested virtualization enabled.  
This is primarily because Android-x86 is not officially compatible with cloud platforms yet~\cite{gcp}, and requires a physical Linux machine with KVM enabled. 

\subsection{Exploring Remote Connection Protocols}

\subsubsection*{{\textbf{Lesson Learned \# 4:}}} \textit{{Existing open-source implementations of remote connection protocols, notably VNC and RDP, yield display in landscape orientation by default. 
This makes it difficult to control the remote device from the smartphone. 
In contrast, WebRTC is a secure alternative that offers screen sharing in portrait mode.}} 

With the server side set up, the next task was to remotely connect the server and the client such that there was no direct delivery of messages to the client device.  
We noticed that Microsoft's Remote Desktop Protocol (RDP)~\cite{rdp} and Virtual Network Computing (VNC)~\cite{vnc} are the two most widely-used remote connection protocols that share the server's screen at the client end, and let the user control the remote server.  
However, existing implementations of RDP and VNC are mostly designed for desktops and always yield display in landscape view. 
The remote landscape view was user-unfriendly when seen on the user's smartphone.  
We attempted changing the screen orientation at the client and server; however, the shared screen remained in landscape mode irrespective of the client's orientation.  
Besides these remote connection protocols, WebRTC provides screen sharing capability in web and native applications. 
As WebRTC offers screen sharing in portrait mode, we combined WebRTC's API with the React web application to provide screen sharing and remote control functionality. 

\subsection{Exploring Remote Connection Applications}

\subsubsection*{{\textbf{Lesson Learned \# 5:}}} \textit{{Unlike computer-to-computer and mobile-to-computer screen sharing applications, there are only a few mobile-to-mobile screen sharing applications available; however, most of them are
costly, work only in the local network, lack remote control functionality, and do not provide unattended access to the server.}} 

We also tested major screen sharing and remote control applications available on the Google Play Store and Apple App Store to see if any application can outperform WebRTC. 
We noticed that most available remote control applications, such as Remote Chrome Desktop~\cite{crd}, RealVNC~\cite{vncviewer}, SplashTop~\cite{splashtop}, ISL Lite~\cite{isllite}, Zoho~\cite{zoho}, and AnyScreen \cite{anyscreen}, required either the server or the client device to be a desktop computer.
In our case, both the client and server run the mobile OS; hence, these applications could not solve the screen orientation issue.

\begin{table}[t]
\centering
\caption{Android based mobile-to-mobile screen sharing utilities and their limitations}
\label{tab:remote}
%\resizebox{\columnwidth}{!}{%
\begin{tabular}{l|l|l} 
\hline
\# & \multicolumn{1}{c|}{\textbf{Solution}} & \multicolumn{1}{c}{\textbf{Limitations}}                                                                  \\ 
\hline
1  & Skype~\cite{skype}                                  & No remote control functionality,                                                                                  \\ 
\hline
2  & ScreenTalk~\cite{screentalk}                             & \begin{tabular}[c]{@{}l@{}}Sluggish performance, \\No remote control functionality,\end{tabular}                                                                             \\ 
\hline
3  & Inkwire~\cite{inkwire}                                & \begin{tabular}[c]{@{}l@{}}Sluggish performance, \\No remote control functionality,\end{tabular}                                                                           \\ 
\hline
4  & Scrcpy~\cite{scrcpy}                                 & \begin{tabular}[c]{@{}l@{}}Requires physical connection to the \\mirrored screen,\end{tabular}                                                                           \\  
\hline
5  & Vysor~\cite{vysor}                                  & \begin{tabular}[c]{@{}l@{}}Requires physical connection to the \\mirrored screen,\end{tabular}                                                                    \\ 
\hline
6  & DroidVNC-ng~\cite{droidng}                               & Unstable (crashes/ does not connect),                                                                                          \\ 
\hline
7 & RemoDroid~\cite{remodroid}                              & \begin{tabular}[c]{@{}l@{}}Worked only in local network, \\Slow, Requires rooting the phone,\end{tabular}  \\

\hline
8 & Airpower~\cite{airpower}                         &  \begin{tabular}[c]{@{}l@{}}Worked only in local network,\\ Sluggish performance,\end{tabular}                                                                           \\ 
\hline
9  & TeamViewer~\cite{teamviewer}                             & Very costly license,                                                                                         \\ 
\hline
10  & Anydesk~\cite{anydesk}                                & Costly license,                                                                                              \\ 
\hline
11  & AirDroid~\cite{airdroid}                               & Free version had limited usage quota.                                                                  \\ 
\hline
\end{tabular}
\end{table}
We, therefore, specifically looked for mobile-to-mobile remote access solutions (Table~\ref{tab:remote}).
While Skype~\cite{skype}, ScreenTalk~\cite{screentalk} and Inkwire~\cite{inkwire} provided high-quality mobile-to-mobile screen sharing, these technologies lacked the required remote control functionality.
Furthermore, applications such as Scrcpy~\cite{scrcpy} and Vysor~\cite{vysor} required physical (USB) access to the smartphone to mirror the screen.  
This was impractical in our case, where the server needs to be a standalone remote device that is always accessible by the user. 
In our search, we found out DroidVNC-ng~\cite{droidng}, RemoDroid~\cite{remodroid} and AirpowerMirror~\cite{airpower} provided both screen mirroring and remote control functionality.
However, Droid-ng and RemoDroid were unstable, and the connection was frequently terminated.
Also, RemoDroid and AirpowerMirror had high latency of up to 8 seconds and could only work within the local network. 
Since our basic requirement was remote user access, we disregarded these options.

Professional proprietary solutions such as TeamViewer~\cite{teamviewer}, AnyDesk~\cite{anydesk} and AirDroid~\cite{airdroid} also meet our requirement of screen sharing and remote access. 
Unfortunately, TeamViewer and AnyDesk had costly licenses, which made it impractical to run the proposed prototype for even a single user. 
In contrast, AirDroid was cost-effective (\$3 a month) and provided high-quality screen mirroring and remote control functionality over the Internet.  
However, as AirDroid is a proprietary application, the user does not have full control over the mirrored data that also passes through the AirDroid infrastructure. 
Thus, we preferred using the WebRTC-based screen sharing service over remote control applications.

\subsection{Open-Ended Security and Usability Issues}

Our framework aims to ensure that the zero-click infection is confined only to the targeted application in case of an attack, while the client device and other isolated remote applications remain safeguarded.  
Below, we highlight the missing components that we deem necessary to achieve better security guarantees against zero-click attacks, and enhance the user's experience of a virtual smartphone.

\paragraph{Security:}

Since our implementation uses COTS components, any vulnerability on the remote server can infect the client device, and invade the user's privacy.
For utmost protection, the server should remain updated to the latest version. 
In practice, any installed chat app or SMS Gateway can also be malicious or compromised.
However, in such a case, only that specific instance will be compromised, and other isolated apps or the client device will not be affected.

Moreover, as iMessages can only be exchanged between Apple devices, shifting iMessage to the server requires a dedicated iOS device which is not a cost-effective solution.
On the contrary, allowing iMessages on the client device can compromise the security of the client device. 
Our framework makes a trade-off between the client device's safety and additional features that iMessage provides over SMS (e.g., end-to-end encrypted chat, screen effects, etc.).
Essentially, we shifted SMS onto the remote server; hence, now iMessages are received as regular SMS on the server and securely mirrored on the client device.  

In addition, the MVT~\cite{mvt} tool can be used to scan the dump of containerized instances every few days and check if any instance is infected with zero-click exploits. 
However, the detection of IoCs is dependent on how frequently the MVT repository is updated. 

\paragraph{Usability:}
With regards to usability, the current off-the-shelf implementation we evaluated to gain experience about zero-click solution requirements does not have a notification mechanism, i.e., the user has to manually check for new messages. 
A possible solution would be to develop an application that reads the status of other applications on the server, and notifies the client device whenever a notification is received. 
However, this is not possible without rooting/jailbreaking the phone, as Android OS and iOS run each application in the sandbox, and have limited content sharing across applications. 
However, rooting the device makes it vulnerable to many other attacks.
Hence, like secure government systems that require the user to check messages manually, there is a trade-off taken between security and usability. 

In contrast to Apple's Lockdown mode that severely restricts the download of attachments, e.g., files on the phone, our setup allows users to download, view and access files remotely.
However, as the remote instances are periodically reset, any received file cannot be saved for future use. The user has to decide the frequency of resetting the server as per his needs. 

Beside text messages, images and files, the virtual smartphone can also receive videos and audio messages. However, since the server's messages are displayed through screen sharing on the user's smartphone, the user can only view the video but not listen to the audio.  
Hence, there would also be a need to effectively relay the audio of communicating parties so that the user can listen to audio messages and secure calls. 

\section{Conclusion} \label{conclusion}

Zero-click attacks are incredibly difficult to detect and pose a pervasive threat to the privacy and security of smartphone users around the world. 
This paper reports on our experiences and the lessons learned while attempting to design and implement a secure framework with standard, off-the-shelf components to protect and limit the impact of a potential zero-click attack on smartphones.
We enumerated several design requirements and presented our envisioned security architecture that shifted the attack surface from the user's device to a sandboxed virtual smartphone ecosystem where sensitive applications run in isolation. 
The users interacted with remote applications using screen sharing and remote access service to prevent the zero-click exploit from accessing the device. 
We demonstrated that it is indeed feasible to practically build such a secure framework using COTS components by accessing chat applications running remotely in containerized Android Emulators using WebRTC-based screen sharing. 
We tested the performance and usability of our implementation through an IRB-exempted research study.
Finally, we highlighted the missing components necessary to achieve security against zero-click attacks and improve users' experience. 
We hope this research will help others build effective remote and isolated systems against zero-click smartphone spyware.

{\normalsize \bibliographystyle{unsrt} %plain
\bibliography{paper}}

\section*{A: Appendix}  
\subsection*{Survey Form for User Study}
Zero-click spyware exploits zero-day (unknown) vulnerabilities in chat applications to stealthily compromise a smartphone without any user interaction.

Our solution curtails the risk of zero-click attacks by running chat apps in isolation at a remote instance, and allows the user to access chat apps via a screen sharing and remote control utility. In this way, if the attacker compromises one chat app, the targeted device and other chat apps remain unaffected. 

You can access the server by visiting https://34.xxx.xxx.xxx from your web browser. You are requested to use the system and provide your valuable feedback. We appreciate your time, effort and willingness to take part in this survey. 

\subsection*{A.1   Background Knowledge}
Have you had any idea about zero-click attacks prior to starting this user study?
\begin{itemize}
    \item Yes,
    \item No.
\end{itemize}

\subsection*{A.2   Remote Connection}
When connecting to the remote server, was the server accessible (in the first attempt) and the connection stable (i.e., it did not terminate after some time)?
\begin{itemize}
    \item Accessible and stable - I seamlessly connected to the remote end,
    \item Accessible but unstable - I connected for a while but the connection terminated later on,
    \item Inaccessible - I could not connect at all.
\end{itemize}

\subsection*{A.3  System's Usability}
Considering the need of a secure solution, do you find the system usable? 
\begin{itemize}
    \item Usable: user-friendly, text is readable, lag is acceptable (considering the immediate need of the solution and its benefits),
    \item Partially usable: user-friendly, text is readable, but experienced excessive lag,
    \item Not usable: Not user-friendly, text is not readable, and experienced excessive lag.
\end{itemize}

\subsection*{A.4  Suggestions}
What can we do to improve your experience (i.e., with regard to the system's performance and usability)?
\end{document}